\documentclass[twocolumn,showpacs,preprintnumbers,amsmath,amssymb,showkeys]{revtex4}

\usepackage{graphicx}
\usepackage{dcolumn}
\usepackage{bm}

\begin{document}


\title{Sequence composition and environment effects on residue fluctuations in protein structures}

\author{Anatoly M. Ruvinsky}
 \email{ruvinsky@ku.edu}

\author{Ilya A. Vakser$^*$}
\email{vakser@ku.edu}
\affiliation{
$^*$Center for Bioinformatics and $^{\dagger}$Department of Molecular Biosciences, The University of Kansas, Lawrence, Kansas 66047}

\date{\today}

\begin{abstract}

The spectrum and scale of fluctuations in protein structures affect the range of cell phenomena, including stability of protein structures or their fragments, allosteric transitions and energy transfer. The study presents a statistical-thermodynamic analysis of relationship between the sequence composition and the distribution of residue fluctuations in protein-protein complexes. A one-node-per-residue elastic network model accounting for the nonhomogeneous protein mass distribution and the inter-atomic interactions through the renormalized inter-residue potential is developed.
Two factors, a protein mass distribution and a residue environment, were found to determine the scale of residue fluctuations. Surface residues undergo larger fluctuations than core residues, showing agreement with experimental observations.  
Ranking residues over the normalized scale of fluctuations yields a distinct classification of amino acids into three groups: (i) highly fluctuating - Gly, Ala, Ser, Pro and Asp, (ii) moderately fluctuating - Thr, Asn, Gln, Lys, Glu, Arg, Val and Cys (iii) weakly fluctuating - Ile, Leu, Met, Phe, Tyr, Trp and His. The structural instability in proteins possibly
relates to the high content of the highly fluctuating residues and a deficiency of the weakly fluctuating residues
in irregular secondary structure elements (loops), chameleon sequences and disordered proteins. Strong correlation between residue fluctuations and the sequence composition of protein loops supports this hypothesis. Comparing fluctuations of binding site residues (interface residues) with other surface residues shows that, on average, the interface is more rigid than the rest of the protein surface and Gly, Ala, Ser, Cys, Leu and Trp have a propensity to form more stable docking patches on the interface. The findings have broad implications for understanding mechanisms of protein association and stability of protein structures.

\end{abstract} 

\maketitle



\section{Introduction}

A remarkable difference between sequence compositions of regular and irregular secondary structure elements of proteins has been  attracting considerable attention for more than 30 years \cite{ML,JFLeszczynski,Richardson,KTO,Kwasigroch,Costantini}.
Amino-acid composition profiles revealed that the irregular regions (protein loops) are enriched in Gly, Pro, Ser and Asp. The regular regions ($\alpha-\text{helices}$ and $\beta-\text{strands}$) contain less of these amino acids. Helices are enriched in Leu, Ala, Glu and Gln, and  $\beta-\text{strands}$ are enriched in Val, Ile, Phe and Tyr. Amino-acid compositions of protein interfaces has been analyzed \cite{Glaser,Keskin,BuyongMa05132003,Ofran2003377}. Despite the extensive use of the statistics in almost all aspects of protein modeling (e.g. in computational algorithms for the secondary structure assignments (see \cite{Rost} for the review);
in knowledge-based approaches to predict protein structures \cite{Tanaka,Miya,Simons,Skolnick,Buchete,Summa}, in receptor-ligand docking \cite{Zh,Ruvinsky,Dars}) the understanding of mechanisms that form amino acids propensities is still incomplete and poses a challenge for researchers in physics and biology. Recent discoveries of chameleon sequences, that undergo helix-sheet transitions \cite{Kim,MMeze,Wei-Zen,Igor,Tidow,Guo}, and intrinsically disordered proteins or fragments, that undergo order-disorder transitions \cite{Wright1999321,Dunker200126,Zha}, have added interest to the problem. Studying the distribution, the scale and features of structural and thermal fluctuations in proteins is one way to tackle this puzzle.  

Protein functionality, encoded into the sequence, is based on a dual ability of proteins to sustain and change their structures \cite{Petsko}. The relationship has different degrees of sensitivity to the location and the scale of changes of protein structures (e.g. $\text{CH}_3$ group rotations, conversions of side-chain rotamers, cis-trans isomerization of proline or domain shifts). Last ten years demonstrated increasing popularity of low-resolution or coarse-grained models in conjunction with harmonic potentials, called elastic network models (ENM), for deciphering and modeling various large-scale structural changes (e.g., allosteric changes in protein structures \cite{Xu2003153,Miyashita,Wentt,Thirumalai}, structural changes on transition  pathways \cite{Miyashita,Hinsen,Kim2002151,Maragakis2005807,Tama,Hummer,Voth}, global conformational changes upon protein-protein binding \cite{Dror,Dina,Sara}). Other applications of these models include the analysis of Debye-Waller factors of $C_{\alpha}$ atoms \cite{Xu2003153,Bahar1997173,ANM,Thorpe,Kondras,Smith,KHinsen} and protein docking \cite{Dina,Zacharias}.     
 
Two types of ENMs are widely used: homogeneous and nonhomogeneous models. A homogeneous ENM is a network of nodes represented by $C_{\alpha}$ atoms and connected by Hooke springs if the distance between nodes is less than a cutoff radius \cite{Xu2003153,Miyashita,Tama,Hummer,Voth,Dror,Sara,Bahar1997173,ANM}. All network nodes are assigned an equal mass that smoothes protein mass density. The homogeneous ENM has two parameters only, the cutoff radius and the spring force constant. Nonhomogeneous ENMs introduce structural and interaction inhomogeneity by assigning residue masses to the network nodes represented by $C_{\alpha}$ atoms \cite{Smith} or by assigning distance- or residue type-dependent force constants to interacting nodes \cite{Thirumalai,Hinsen,Maragakis2005807,Kondras,Smith,KHinsen,Zacharias}. The effect of protein sequence variations on the spring force constants 
has been considered recently \cite{Wentt}.  
Double-well ENMs are used to model large-scale conformational transition pathways  \cite{Maragakis2005807,Hummer,Voth,Koga}. Merging residues into rigid blocks is used to consider properties of large macromolecules within ENM of a lower resolution \cite{Miyashita,Tama,Thorpe,Blocks}. Less ``extreme'' coarse graining keeps three translational degrees of freedom of $C_{\alpha}-\text{based}$ nodes and degrees of freedom of bond angles and dihedrals (see \cite{Koga,McCammon,Miri}).
 
In the context of nonhomogeneous ENMs, we present a novel
method to account for the protein mass distribution and inter-atomic contacts within the coarse-grained model. We move network nodes from $C_{\alpha}$ atoms to the centers of mass of protein residues to bring in the effects of side chains into the model. We derive a modified Tirion-like potential \cite{Tirion} to bring in structural details of the atomic level and put forward a statistical-thermodynamic formalism to calculate residue fluctuations of a set of protein complexes \cite{Gao}. We show that the scale of residue fluctuations increases from the inside to the protein surface, showing agreement with the Frauenfelder-Petsko-Tsernoglou model \cite{Hans}. 
We suggest a classification of protein residues based on the normalized scale of fluctuations and discuss how the scale of fluctuations correlates with amino acid propensities in secondary structure elements, chameleon sequences and disordered fragments. Fluctuations of binding site residues (interface residues) are compared with other surface residues. The tendency of some residues to form more stable docking patches on the interface is discussed as well as the role of loops at early stages of protein thermal denaturation.

\section{Model}

A modified nonhomogeneous ENM is used in calculations. Network nodes are placed in the centers of mass of protein residues and residue masses are assigned to the corresponding network nodes. The following is a description of a formalism to consistently transform the inter-atomic protein energy landscape into the inter-residue landscape. As a result, we obtain a modified inter-residue harmonic potential with a spring force constant proportional to the number of inter-atomic contacts between residues (see Eq~(\ref{eq:02}) below). 

The interaction energy between protein residues $i$ and $k$ is
\begin{equation}
U_{ik}({\bm R}_i-{\bm R}_k)=\sum\limits_{\alpha,\beta} U_{\alpha\beta}({\bm R}_i+{\bm u}_{\alpha}^i-{\bm R}_k-{\bm u}^k_{\beta}),
\label{eq:01}
\end{equation}
where ${\bm R}_{i,k}$  are radius vectors of the centers of mass of residues $i$ and $k$, 
${\bm u}_{\alpha,\beta}^{i,k}$ are the radius vectors of atoms $\alpha$ and $\beta$ relative to the centers of mass of the residues $i$ and $k$ accordingly. The sum in Eq.~(\ref{eq:01}) runs over all pairs of atoms separated by a distance less then the interaction cutoff. We use the cutoff of $14\text{\AA}$ that assures a tolerable level of the cutoff-related ruggedness of the energy landscape \cite{Ruvinsky02232009,RAMV}. 
Introducing a residue-residue potential, one can rewrite Eq.~(\ref{eq:01}) as $U_{ik}({\bm R}_i-{\bm R}_k)=N_{ik}V({\bm R}_i-{\bm R}_k)$, where $N_{ik}$ is a number of inter-atomic interactions between residues $i$ and $k$ and  $V$ is the coarse-grained or inter-residue potential {\it per se}. Assuming that 
inter-residue interactions are in equilibrium in a native protein and using a Lennard-Jones form of the inter-residue potential, we can expand $V({\bm R}_i-{\bm R}_k)$ in Taylor series of deviations $R_{ik}-R_{ik}^0$ of the inter-residue distance $R_{ik}=|{\bm R}_i-{\bm R}_k|$ from its equilibrium $R_{ij}^0$. Expanding to the second order in $R_{ij}-R_{ij}^0$ yields 
\begin{equation}
\label{pot}
U_{ik}({\bm R}_i-{\bm R}_k)=-N_{ik}\varepsilon +36N_{ik}\varepsilon \left(\frac{R_{ik}-R_{ik}^0}{R_{ik}^0}\right)^2,
\end{equation}
where $\varepsilon$ is the depth of the Lennard-Jones potential. Eq.~\ref{pot} shows that inter-residue interactions are proportional to the number of inter-atomic interactions and decrease with the increase of the inter-residue distance as $1/(R_{ik}^0)^2$. 

Since ${\bm R}_{i,k}={\bm R}_{i,k}^0+{\bm r}_{i,k}$, we obtain
\begin{equation}
U_{ik}({\bm r}_i-{\bm r}_k,{\bm R}_{ik}^0)=-\varepsilon N_{ik} +36\varepsilon \frac{N_{ik}}{(R_{ik}^0)^2}\left(\frac{{\bm R}_{ik}^0}{R_{ik}^0}({\bm r}_{i}-{\bm r}_{k})\right)^2,
\label{eq:02}
\end{equation}   
where ${\bm r}_{i,k}$  are the deviations of the residue centers of mass from its equilibrium position. 
The main difference between Eq~(\ref{eq:02}) and Tirion-like potentials \cite{Tirion} used in nonhomogeneous ENMs is the factor $N_{ik}$ which introduces the distribution of inter-atomic interactions into the coarse-grained model. In other words, 
the change of the protein model resolution from the atomic to the residue level results in the appearance of this factor in the inter-residue potential. 

The protein Lagrangian   
\begin{equation}
{\cal L}=\sum\limits_{i,k=1}^N \frac{m_i}{2}(\dot {\bm r}_i)^2-U_{ik}({\bm r}_i-{\bm r}_k,{\bm R}_{ik}^0)
\label{eq:03}
\end{equation} 
derives the following $3N$ equations of motions 
\begin{equation}
m_i\ddot {\bm r}_i=-\sum\limits^{N}_{k=1} C_{ik} \left({\bm\alpha}_{ik}({\bm r}_i-{\bm r}_k)\right){\bm\alpha}_{ik},
\label{}
\end{equation} 
where $m_i$ is the mass of the residue $i$, ${\bm\alpha}_{ik}={\bm R}_{ik}^0/R_{ik}^0$, $C_{ik}=72\varepsilon N_{ik}/(R_{ik}^0)^2$ and $N$ is the number of protein residues. As usual, we seek an oscillatory solution of the form ${\bm r}_k={\bm A}_k\exp(i\omega t)$, where $A_k$ are some amplitude factors to be determined. The substitution of the trial solution into the equations of motions leads to the eigenvalue problem $({\bm H}-\omega^2{\bm I}){\bm A}=0$, where $A=\{A_1^x,A_1^y,A_1^z,A_2^x,A_2^y,\dots\}$ is a $3N$ column vector of the amplitude factors, ${\bm I}$ is a $3N\times 3N$ unit matrix, ${\bm H}$ is a $3N\times 3N$ matrix composed of $3\times 3$ super elements
\begin{equation}
H_{ik}(i\ne k)=\left[
\begin{array}{ccc}
h_{ik}^{xx}  & h_{ik}^{yx} & h_{ik}^{zx} \\
h_{ik}^{xy}  & h_{ik}^{yy} & h_{ik}^{zy} \\
h_{ik}^{xz}  & h_{ik}^{yz} & h_{ik}^{zz} \\

\end{array}
\right], H_{ii}=-\sum\limits_k^\prime H_{ik},
\label{eq:04}
\end{equation}
where $h_{ik}^{ab}=-C_{ik}\alpha_{ik}^{a}\alpha_{ik}^{b}/m_i$ and the upper indexes $a, b$ stand for $x,y,z$ projections of 
the vector ${\bm\alpha}_{ik}$.

The prime in sums
over $k$ in Eqs.~(\ref{eq:04}) means that a term $i=k$ is not accounted for. We use our program AH ({\bf A}nalyzer of {\bf H}armonics) to find protein eigenfrequencies 
$\{\omega\}$ and normalized eigenvectors. The {\it k}th oscillation can be written in the form
\begin{equation}
x_k=\sum\limits_{i=1}^{3N-6}G_{ki}c_i\exp(\omega_i t)=\sum\limits_{i=1}^{3N-6}G_{ki}\Theta_i,
\label{eq:05}
\end{equation}
where $\Theta_i=Re[c_i\exp(\omega_it)]$ is the the so called normal coordinate, $Re$ stands for ``real part of,'' $c_i$ is a constant determined by initial conditions,
columns of the matrix ${\bm G}$ are the normalized eigenvectors. The normal modes are described by 
\begin{equation}
{\cal H}=\sum\limits_{i=1}^{3N-6} \frac{M_i}{2}\left(\dot \Theta_i^2+\omega_i^2\Theta_i^2\right),
\label{eq:06}
\end{equation}
where $M_i=\sum\limits_{k=1}^{3N-6} m_kG_{ki}^2$ is the effective mass of the {\it i}th normal mode \cite{Landau}. 
Note that for a homogeneous ENM, $m_i$ is a constant equal to some parameter $m$ and, therefore, all modes will have equal effective masses: $M_i=\sum\limits_{k=1}^{3N-6} mG_{ki}^2=m$.  

The mean-square fluctuation of the \textit{k}th residue along the coordinate axis $x$ is
$<x_{k_x}^2>=\sum\limits_{ij}G_{k_xi}G_{k_xj}<\Theta_i\Theta_j>$, where the angular brackets denote a Boltzmann average with the Hamiltonian (\ref{eq:06}) over the normal modes, $k_{x,y,z}$ are the numbers of degrees of freedom associated with the residue center of mass oscillations along the coordinate axes $x,y,z$. Boltzmann averaging of pair products $<\Theta_i\Theta_j>$ of normal coordinates yields $<\Theta_i\Theta_j>=\delta_{ij}Tk_B/(M_i\omega_i^2)$, where $T$ is the temperature, $k_B$ is the Boltzmann constant and $\delta_{ij}$ is the Kronecker delta ($\delta_{ij}=1$ if $i=j$ and $\delta_{ij}=0$ if $i\ne j$). The total mean-square fluctuation of the \textit{k}th residue has the form
\begin{equation}
<{\bm r}_k^2>=Tk_B\sum\limits_{i=1}^{3N-6} \frac{G_{k_xi}^2+G_{k_yi}^2+G_{k_zi}^2}{M_i\omega_i^2}
\label{eq:07}
\end{equation} 
It is important to note that the residue fluctuation, derived in Eq (\ref{eq:07}), shows nonlocal dependence on the mass distribution in a protein. This effect totally disappears in the framework of a homogeneous ENM.   

Removing the effect of the parameter $\varepsilon$ on residue fluctuations, we introduce a mobility ratio (MR) of the {\it k}th residue in the form 
\begin{equation}
{\cal R}_k=\frac{<{\bm r}_k^2>}{{\bm r}^2_{av}},
\end{equation} where 
${\bm r}^2_{av}=\sum\limits_{k=1}^N<{\bm r}_k^2>/N$ 
is the averaged mean-square fluctuation in a protein. 

We computed the mobility ratios for each of the protein residues in 184 proteins from the 92 non-obligate protein-protein complexes selected from a docking benchmark set \cite{Gao}. For each of the proteins MRs were grouped in twenty groups according to names of standard amino acids and twenty average MRs were computed. The obtained values were averaged over the set of 184 protein structures.  
Figures 1-3 show mean MRs and standard deviations of the mean.

\section{Results}

\begin{figure}[]
\includegraphics[scale=0.88]{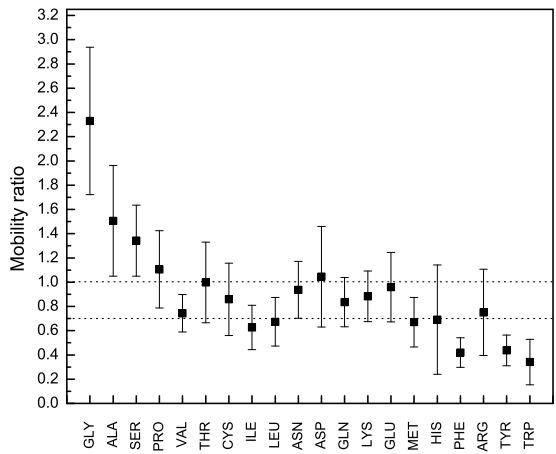}
\caption{The mobility ratios of protein residues arranged in the order of increasing mass.}
\label{fig1}
\end{figure}

The results show that large equilibrium fluctuations (${\cal R}\geq 1$) of protein structures are associated with the oscillations of the center of mass of Gly, Ala, Ser, Pro and Asp (Group I) which are the most lightweight residues with the exception of Asp (Fig.~\ref{fig1}).  Modest fluctuations (${\cal R}=0.7\div 1.0$; Group II) are associated with six polar residues (Thr, Asn, Gln, Lys, Glu, Arg) and two nonpolar residues (Val, Cys).  The small fluctuations (${\cal R}=0.3\div 0.7$; Group III) are associated with six nonpolar residues (Ile, Leu, Met, Phe, Trp) and polar residues His and Tyr. It is interesting to note that, with regards to hydrophilicity, groups I, II and III can be characterized as mixed, mostly polar and mostly nonpolar.  

\begin{figure}[]
\includegraphics[scale=0.88]{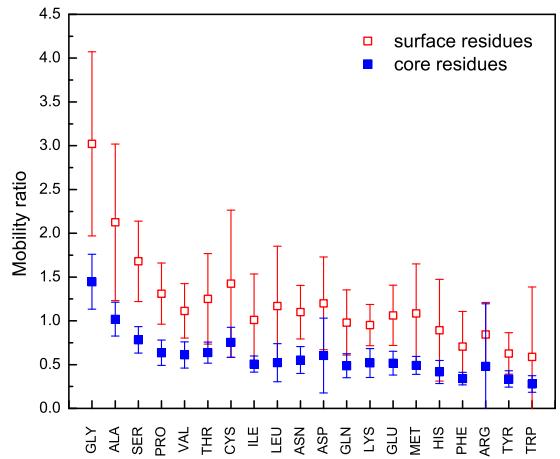}
\caption{The mobility ratios of surface and core protein residues. Surface (core) residues are defined here as those residues which have relative solvent accessible surface area higher(lower) than $25\%$.}
\label{fig2}
\end{figure}

Analysis of the scale of fluctuations of surface and core residues shows that on average all surface residues demonstrate larger fluctuations than core residues (Fig.~\ref{fig2}). Surface (core) residues are defined here as those residues which have relative solvent accessible surface area higher(lower) than $25\%$ and are identified using NACCESS \cite{Hub}. The difference is readily explained by the difference in numbers of nearest neighbors of surface and core residues (the environment effect). In comparison with core residues, surface residues have less nearest neighbors \cite{Packing}. Therefore, they are less restricted and experience larger fluctuations. First reports of this effect go back to crystallographic studies of myoglobin \cite{Hans,Greg} and 
lysozyme \cite{Artem}. It has been showed that atomic mean-square displacements increase from the protein core to the protein surface.  Frauenfelder {\it et al} \cite{Hans} suggested that, in general, proteins have a condensed core and a semi-liquid surface.

The same environment effect appears as a small root mean-squared deviation between bound and unbound states of pocket side chains \cite{Li2004781} or as a decreased number of rotamers allowable for buried amino acids in comparison with surface amino acids \cite{Deep,Smith20051077}. This also clears up a seemingly striking difference in hydrophilicity found between residues of groups II and III. Indeed, amino acid residues are distributed nonhomogeneously in proteins. Polar residues prefer surface positions, but unpolar residues are more often found in a protein core. That is why the mostly polar group II demonstrates higher mobility ratios than the mostly unpolar group III. On the other side, high mobility ratios of nonpolar residues Gly and Ala suggest that the environment effect is not the only factor. The amplitude of fluctuations is inversely proportional to the effective amino acid masses (see Eq.~\ref{eq:07}). As a result, the largest fluctuations are accosiated with Gly and Ala, the most lightweight residues, but the smallest fluctuations are accosiated with Tyr and Trp, the most heavy residues (Fig.~\ref{fig1}). 

\begin{figure}[]
\includegraphics[scale=0.88]{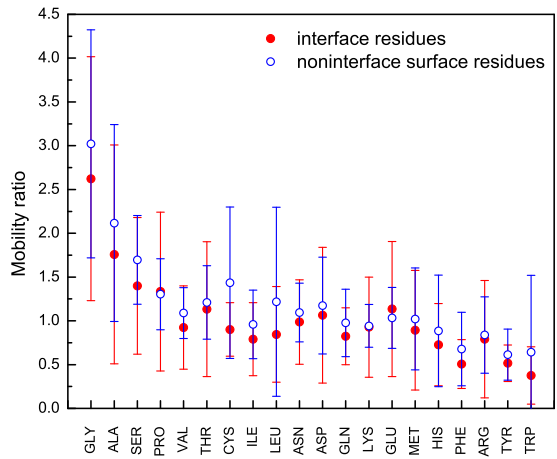}
\caption{The mobility ratios of interface and non-interface surface protein residues.}
\label{fig3} 
\end{figure}

Comparing fluctuations of binding site (interface) residues with other surface residues, we found that although, on average, 
interface is less mobile than the rest of the protein surface (Fig.~\ref{fig3}), the noticeable difference (${\cal R}^{sur}_j-{\cal R}_j^{int}>0.25$, ${\cal R}_j^{int,sur}$ is the mobility ratio of the interface or other surface residue j) relates to Gly, Ala, Ser, Cys, Leu and Trp. Four of these residues (Gly, Ala, Leu and Ser) are the most common residues at protein interfaces, and residues Cys and Trp are the most infrequent interface residues \cite{Glaser,Keskin}. The most conserved interface residue Trp \cite{BuyongMa05132003} also
is the most stable one (see Fig.~\ref{fig3}). Two other highly conserved interface residues (Met and Phe) \cite{BuyongMa05132003} demonstrate decreased mobility in binding cites to a lesser extent. Note that the difference between binding cites and the rest of the protein surface relates mainly to fluctuations of the nonpolar residues with the exception of Ser, a polar residue. 
These results are in agreement with the experimental observation of reduced fluctuations in binding sites of myoglobin \cite{Ben} and bacteriorhodopsin \cite{Reat} in comparison with fluctuations of the rest of macromolecules. Frauenfelder and McMahon \cite{Ben} also noted that the four (Leu29, Phe43, Val68 and Ile107) of the six residues with reduced fluctuations surrounding the oxygen molecule are nonpolar. The two other residues are His64 and His93 (${\cal R}^{sur}_{His}-{\cal R}^{int}_{His}=0.16$).  
The solvent-mediated attraction between nonpolar residues of a receptor and a ligand results in the hydrophobic contribution to binding free energy, which is considered to be one of major factors stabilizing protein-protein complexes \cite{Aflalo,Young,Jones1997133,Tsai}. We suppose that Gly, Ala, Ser, Cys, Leu and Trp form low-mobility surface ``pads'' that   
constitute a ``landing ground'' for binding proteins. 

\begin{figure}[]
\includegraphics[scale=0.88]{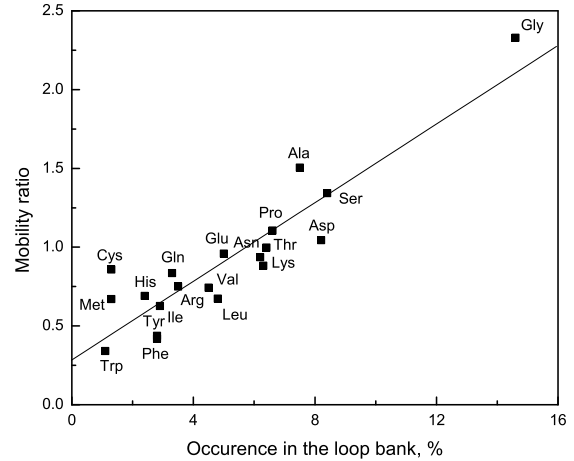}
\caption{The mobility ratios of protein residues against their percentage compositions in protein loops
\cite{Kwasigroch}.}
\label{fig4} 
\end{figure}

The larger ability to fluctuate of Group I residues provides an insight into the inability of sequences abundant in Gly, Ala, Ser, Pro and Asp to fold into regular protein secondary structure elements ($\alpha-\text{helices}$ or $\beta-\text{strands}$). High mobility prevents the formation of long-range order thus contributing to irregular protein secondary structure elements (loops). We computed the correlation coefficient between the mobility ratios and corresponding percentages of amino acid residues in the bank of loops \cite{Kwasigroch} (see Fig.~(\ref{fig4})). The analysis showed significant relation
with $0.9$ correlation coefficient.

We suggest that the same reasoning explains features of amino acid distributions observed in chameleon sequences \cite{Kim,MMeze,Igor,Guo} and disordered proteins  \cite{Dunker200126,Zha}. Indeed, highly and moderately fluctuating amino acid residues (in particular, Gly, Ala, Ser, Glu and Lys) are abundant in disordered and ``dual personality'' protein fragments, whereas the residues with the low mobility ratio (e.g. Tyr, Trp, Phe, Ile) are rarely found there \cite{Dunker200126,Zha}.

Statistics of protein residues in chameleon sequences shows that Ala, Ile, Leu and Val are the most frequent residues in chameleon sequences \cite{MMeze,Guo}. Since only Ala belongs to the Group I of highly fluctuating residues (Fig.~\ref{fig1}), we can hypothesize that an instability driving $\text{helix}\leftrightarrow\text{sheet}$ transitions may often originate at Ala residues if the other highly fluctuating residues are absent. Frequencies of occurrence of Gly and Ser residues increase with the increase of the length of the sequence \cite{MMeze}. Thus, in general, chameleon sequences may have several islands of instability. Exciting these islands locally (e.g., by mutations that change interactions of the islands with the rest of the protein or by ligands bound in the vicinity of the chameleon sequence), one could trigger a $\text{helix}\leftrightarrow\text{sheet}$ transition. Mutations of a chameleon sequence, that change the mobility ratio of a sequence position  significantly, can also provoke such transitions. It has been reported that  a single mutation from Pro to Ala (${\cal R}_{Ala}-{\cal R}_{Pro}=0.4$) converts a $\beta-\text{sheet}$ into an $\alpha-\text{helix}$ \cite{Wei-Zen}. Mutations of two consecutive residues from Phe28Phe29 to Pro28Ile29 (${\cal R}_{Pro}-{\cal R}_{Phe}=0.7$, ${\cal R}_{Ile}-{\cal R}_{Phe}=0.2$) converts an $\alpha-\text{helix}$ into a $\beta-\text{sheet}$\cite{Tidow}.

The mobility ratio derived by Eq~(\ref{eq:07}) increases with the temperature increase. Therefore, we could expect that at the very early stages of protein thermal denaturation amino acid residues of the enhanced ability to fluctuate (Group I) and their structural neighbors will form first seeds of the unfolded phase. Since the majority of Group I amino acids (Gly, Ser, Pro and Asp) shows higher propensities for loops than for helices or sheets \cite{Costantini}, it is possible that the nucleation of the unfolded phase starts on protein loops. Due to the increased ability to fluctuate, Group I residues can be also involved  more often than other residues into equilibrium local folding-unfolding reactions scattered over the protein surface \cite{YBai07141995,Fierz}.  

\section{Conclusions}

The current work focuses on the fundamental relationship between the protein sequence, ability to fluctuate and functionality 
of protein structures. We have considered the relationship within a framework of a novel elastic network model that allows accounting for the distribution of inter-atomic interactions within a coarse-grained approach. The model modifies a commonly used form of the Tirion potential with a spring constant proportional to the number of inter-atomic contacts between residues. 
We demonstrated that two factors, a protein mass distribution and a residue environment, determine the scale of fluctuations. Surface residues undergo larger fluctuations than core residues in agreement with experimental observations \cite{Hans,Greg,Artem}.  
On average, the protein interface is less mobile than the rest of the protein surface and contains low-mobility pads associated mainly with nonpolar residues. We hypothesize that the conformational instability of protein loops, chameleon sequences and disordered proteins relates to the high content of highly mobile residues and the lack of weakly fluctuating residues. The results show high correlation between fluctuations and the sequence composition of protein loops. Analysis of residue fluctuations and their propensities in secondary structure elements allows one to conclude that upon thermal denaturation the nucleation of the unfolded phase proceeds from protein loops. The results 
provide insight into structural fluctuations of proteins 
and facilitate better understanding of protein association mechanisms.

\begin{acknowledgments}
The study was supported by R01 GM074255 grant from NIH.
\end{acknowledgments}


\begin{thebibliography}{1000}


\bibitem{ML}
Levitt M (1978) Conformational preferences of amino acids in globular proteins.
{\it Biochem} 17: 4277-4285.  

\bibitem{JFLeszczynski}
Leszczynski JF and Rose GD (1986) Loops in globular proteins: a novel category of secondary structure.
{\it Science} 234: 849-855. 

\bibitem{Richardson}
Richardson JS and Richardson DC (1988) Amino acid preferences for specific locations at the ends of alpha helices.
{\it Science} 240: 1648-1652.

\bibitem{KTO}
O'Neil KT and DeGrado WF (1990) A thermodynamic scale for the helix-forming tendencies of the commonly occurring amino acids.
{\it Science} 250: 646-651.  

\bibitem{Kwasigroch}
Kwasigroch J-M, Chomilier J and Mornon J-P (1996) A Global Taxonomy of Loops in Globular Proteins. 
{\it J Mol Biol} 259: 855-872.
 
\bibitem{Costantini}
Costantini S, Colonna G and Facchiano AM (2006)
Amino acid propensities for secondary structures are influenced by the protein structural class.
{\it Biochem Biophys Res Comm} 342: 441-451. 

\bibitem{Glaser}
Glaser F, Steinberg DM, Vakser IA and Ben-Tal N (2001) Residue frequencies and pairing preferences at protein-protein interfaces. 
{\it Proteins} 43: 89-102.

\bibitem{Keskin}
Keskin O, Bahar I, Badretdinov AY, Ptitsyn OB and Jernigan RL (1998) Empirical solvent-mediated potentials hold for both intra-molecular and inter-molecular inter-residue interactions. {\it Prot Sci} 7: 2578 - 2586. 

\bibitem{BuyongMa05132003}
Ma B, Elkayam T, Wolfson H and Nussinov R (2003) Protein-protein interactions: structurally conserved residues distinguish between binding sites and exposed protein surfaces. {\it Proc Natl Acad Sci USA} 100: 5772-5777.  

\bibitem{Ofran2003377}
Ofran Y and Rost B (2003) Analysing six types of protein-protein interfaces. {\it J Mol Biol} 325: 377-387.  

\bibitem{Rost}
Rost B (2001) Review: protein secondary structure prediction continues to rise. {\it J Struct Biol} 134: 204-218.

\bibitem{Tanaka}
Tanaka S and Scheraga HA (1976) Medium- and long-range interaction parameters between amino acids for predicting three-dimensional structures of proteins. {\it Macromol} 9: 945-950.  

\bibitem{Miya}
Miyazawa S and Jernigan RL (1985) Estimation of effective interresidue contact energies from protein crystal structures: quasi-chemical approximation. {\it Macromol} 18: 534-552.

\bibitem{Simons}
Simons KT, Kooperberg C, Huang E and Baker D (1997) Assembly of protein tertiary structures from fragments with similar local sequences using simulated annealing and bayesian scoring functions. {\it J Mol Biol} 268: 209-225.   

\bibitem{Skolnick}
Lu H and Skolnick J (2001) A distance-dependent atomic knowledge-based potential for improved protein structure selection.
{\it Proteins} 44: 223-232.  

\bibitem{Buchete}
Buchete N-V, Straub JE and Thirumalai D (2004) Development of novel statistical potentials for protein fold recognition. {\it Curr 
Opin Struct Biol} 14: 225-232.

\bibitem{Summa}
Summa CM and Levitt M (2007) Near-native structure refinement using in vacuo energy minimization. {\it Proc Natl Acad  Sci USA}
104: 3177-3182.  

\bibitem{Zh}
Zhang C, Liu S, Zhou H and Zhou Y (2004) An accurate, residue-level, pair potential of mean force for folding and binding based on the distance-scaled, ideal-gas reference state. {\it Prot Sci} 13: 400-411.  

\bibitem{Ruvinsky}
Ruvinsky AM and Kozintsev AV (2005) The key role of atom types, reference states, and interaction cutoff radii in the knowledge-based method: New variational approach. {\it Proteins} 58: 845-851.  

\bibitem{Dars}
Chuang G-Y, Kozakov D, Brenke R, Comeau SR and Vajda S (2008) DARS (Decoys As the Reference State) potentials for protein-protein docking. {\it Biophys J} 95: 4217-4227. 

\bibitem{Kim}
Minor DL and Kim PS (1996) Context-dependent secondary structure formation of a designed protein sequence.
{\it Nature} 380: 730-734.

\bibitem{MMeze}
Mezei M (1998) Chameleon sequences in the PDB. {\it Prot Eng} 11: 411-414.

\bibitem{Wei-Zen}
Yang W-Z, Ko T-P, Yuan HS, Corselli L and Johnson RC (1998) Conversion of a beta-strand to analpha-helix induced by a single-site mutation observed in the crystal structure of fis mutant. {\it Prot Sci} 7: 1875-1883.

\bibitem{Igor}
Kuznetsov IB and Rackovsky S (2003) On the properties and sequence context of structurally ambivalent fragments in proteins.
{\it Prot Sci} 12: 2420-2433.

\bibitem{Tidow}
Tidow H, Lauber T, Vitzithum K, Sommerhoff CP, R\"osch R and Marx UC (2004) The solution structure of a chimeric LEKTI domain 
reveals a chameleon sequence. {\it Biochem} 43: 11238-11247. 

\bibitem{Guo}
Guo J-T, Jaromczyk JW and Xu Y (2007) Analysis of chameleon sequences and their implications in biological processes.
{\it Proteins} 67: 548-558.



\bibitem{Wright1999321}
Wright PE and Dyson HJ (1999) Intrinsically unstructured proteins: re-assessing the protein structure-function paradigm. {\it J Mol Biol} 293: 321-331. 

\bibitem{Dunker200126}
Dunker AK, Lawson JD, Brown CJ, Williams RM, Romero P, Oh JS, Oldfield CJ, Campen AM, Ratliff CM, Hipps KW, Ausio J,
Nissen MS, Reeves R, Kang C, Kissinger CR, Bailey RW, Griswold MD, Chiu W, Garner EC and Obradovic Z (2001) Intrinsically disordered protein. {\it J Mol Graph Model} 19: 26-59. 

\bibitem{Zha}
Zhang Y, Stec B and Godzik A (2007) Between order and disorder in protein structures: analysis of "dual personality" fragments in proteins. {\it Structure} 15: 1141-1147. 

\bibitem{Petsko}
Petsko GA and Ringe D (1984) Fluctuations in protein structure from X-ray diffraction. {\it Ann Rev Biophys Bioeng} 
13: 331-371.


\bibitem{Xu2003153}
Xu C, Tobi D and Bahar I (2003) Allosteric changes in protein structure computed by a simple mechanical model: hemoglobin T$\leftrightarrow$R2 transition. {\it J Mol Biol} 333: 153-168. 

\bibitem{Miyashita}
Miyashita O, Onuchic JN and Wolynes PG (2003) Nonlinear elasticity, proteinquakes, and the energy landscapes of functional transitions in proteins. {\it Proc Natl Acad Sci USA} 100: 12570-12575.  

\bibitem{Wentt}
Zheng W, Brooks BR and Thirumalai D (2006) Low-frequency normal modes that describe allosteric transitions in biological nanomachines are robust to sequence variations. {\it Proc Natl Acad Sci USA} 103: 7664-7669.

\bibitem{Thirumalai}
Zheng W and Thirumalai D (2009) Coupling between normal modes drives protein conformational dynamics: illustrations using allosteric transitions in myosin II. {\it Biophys J} 96: 2128-2137.  

\bibitem{Hinsen}
Hinsen K (1998) Analysis of domain motions by approximate normal mode calculations. {\it Proteins} 33: 417-429.

\bibitem{Kim2002151}
Kim MK, Chirikjian GS and Jernigan RL (2002) Elastic models of conformational transitions in macromolecules. {\it J Mol Graph Model} 21: 151-160. 

\bibitem{Maragakis2005807}
Maragakis P and Karplus M (2005) Large amplitude conformational change in proteins explored with a plastic network model: adenylate kinase. {\it J Mol Biol} 352: 807-822. 

\bibitem{Tama}
Tama F and Brooks CL,III (2005) Diversity and identity of mechanical properties of icosahedral viral capsids studied with 
elastic network normal mode analysis. {\it J Mol Biol} 345: 299-314.

\bibitem{Hummer}
Zheng W, Brooks BR and Hummer G (2007) Protein conformational transitions explored by mixed elastic network models. {\it Proteins}
69: 43-57.  

\bibitem{Voth}
Chu J-W and Voth GA (2007) Coarse-grained free energy functions for studying protein conformational changes: a double-well network model. {\it Biophys J} 93: 3860-3871.  

\bibitem{Dror}
Tobi D and Bahar I (2005) Structural changes involved in protein binding correlate with intrinsic motions of proteins in the unbound state. {\it Proc Natl Acad Sci USA} 102: 18908-18913.

\bibitem{Dina}
Schneidman-Duhovny D, Nussinov R and Wolfson HJ (2007) Automatic prediction of protein interactions with large scale motion. {\it Proteins} 69: 764-773.  
 
\bibitem{Sara}
Dobbins SE, Lesk VI and Sternberg MJE (2008) Insights into protein flexibility: The relationship between normal modes and conformational change upon protein-protein docking. {\it Proc Natl Acad Sci USA} 105: 10390-10395. 


\bibitem{Bahar1997173}
Bahar I, Atilgan AR and Erman B (1997) Direct evaluation of thermal fluctuations in proteins using a single-parameter harmonic potential. {\it Fold Des} 2: 173-181.  

\bibitem{ANM}
Atilgan AR, Durell SR, Jernigan RL, Demirel MC, Keskin O and Bahar I (2001) Anisotropy of fluctuation dynamics of proteins with an elastic network model. {\it Biophys J} 80: 505-515.

\bibitem{Thorpe}
Gohlke H and Thorpe MF (2006) A  natural coarse graining for simulating large biomolecular motion. 
{\it J Mol Biol} 91: 2115-2120.

\bibitem{Kondras}
Kondrashov DA, Cui Q and Jr,Phillips GN (2006) Optimization and evaluation of a coarse-grained model of protein motion using X-ray crystal data. {\it Biophys J} 91: 2760-2767. 

\bibitem{Smith}
Moritsugu K and Smith JC (2007) Coarse-grained biomolecular simulation with REACH: realistic extension algorithm via covariance hessian. {\it Biophys J} 93: 3460-3469. 

\bibitem{KHinsen}
Hinsen K (2008) Structural flexibility in proteins: impact of the crystal environment. {\it Bioinform} 24: 521-528.

\bibitem{Zacharias}
May A and Zacharias M (2008) Energy minimization in low-frequency normal modes to efficiently allow for global flexibility during systematic protein-protein docking. {\it Proteins} 70: 794-809.  

\bibitem{Koga}
Okazaki K, Koga N, Takada S, Onuchic JN. and Wolynes PG (2006) Multiple-basin energy landscapes for large-amplitude conformational motions of proteins: Structure-based molecular dynamics simulations. {\it Proc Natl Acad Sci USA} 103: 118444-11849.

\bibitem{Blocks}
Tama F, Gadea FX, Marques O and Sanejouand Y-H (2000) Building-block approach for determining low-frequency normal modes of macromolecules. {\it Proteins} 41: 1-7.  

\bibitem{McCammon}
Tozzini V, Rocchia W and McCammon AJ (2006) Mapping all-atom models onto one-bead coarse-grained models:  general properties and applications to a minimal polypeptide model. {\it J Chem Theo Comput} 2: 667-673.

\bibitem{Miri}
Mirijanian DT and Voth GA (2008) Unique elastic properties of the spectrin tetramer as revealed by multiscale coarse-grained modeling. {\it Proc Natl Acad Sci USA} 105: 1204-1208.

\bibitem{Tirion}
Tirion MM (1996) Large amplitude elastic motions in proteins from a single-parameter atomic analysis.
{\it Phys Rev Lett} 77: 1905-1908.


\bibitem{Gao}
Gao Y, Douguet D, Tovchigrechko A and Vakser IA (2007) Dockground system of databases for protein recognition studies: unbound structures for docking. {\it Proteins} 69: 845-851.

\bibitem{Ruvinsky02232009}
Ruvinsky AM and Vakser IA (2009) The ruggedness of protein-protein energy landscape and the cutoff for $1/r^n$ potentials.   {\it Bioinform}, 25: 1132-1136.  

\bibitem{RAMV}
Ruvinsky AM and Vakser IA (2008) Interaction cutoff effect on ruggedness of protein-protein energy landscape. {\it Proteins}, 70: 1498-1505

\bibitem{Landau}
Landau LD and Lifshitz EM (1976), Mechanics (Pergamon Press, New York), pp 65-70.

\bibitem{Hub}
Hubbard SJ and Thornton JM (1993) 'NACCESS', Computer Program, Department of
Biochemistry and Molecular Biology, University College London. 

\bibitem{Packing}
Gerstein M and Chothia C (1996) Packing at the protein-water interface. 
{\it Proc Natl Aca Sci USA} 93: 10167-10172.

\bibitem{Hans}
Frauenfelder H, Petsko GA and Tsernoglou D (1979) Temperature-dependent X-ray diffraction as a probe of protein structural dynamics. {\it Nature} 280: 558-563.

\bibitem{Greg}
Frauenfelder H and Petsko GA (1980) Structural dynamics of liganded myoglobin. {\it Biophys J} 32: 465–483.

\bibitem{Artem}
Artymiuk PJ, Blake CCF, Grace DEP, Oatley SJ, Phillips DC and Sternberg MJE (1979) Crystallographic studies of the dynamic properties of lysozyme. {\it Nature} 280: 563-568.  

\bibitem{Li2004781}
Li X, Keskin O, Ma B, Nussinov R and Liang J (2004) Protein-protein interactions: hot spots and structurally conserved residues often locate in complemented pockets that Pre-organized in the unbound states: implications for docking.
{\it J Mol Biol} 344: 781-795. 

\bibitem{Deep}
Rajamani D, Thiel S, Vajda S and Camacho CJ (2004) Anchor residues in protein-protein interactions. 
{\it Proc Natl Aca Sci USA} 101: 11287-11292.

\bibitem{Smith20051077}
Smith GR, Sternberg MJE and Bates PA (2005) The relationship between the flexibility of proteins and their conformational states on forming protein-protein complexes with an application to protein-protein docking.
{\it J Mol Biol} 347: 1077-1101.   

\bibitem{Reat}
R\`{e}at V, Patzelt H, Ferrand M, Pfister C, Oesterhelt D and Zaccai G (1998) Dynamics of different functional parts of bacteriorhodopsin: H-$^2$H labeling and neutron scattering. {\it Proc Natl Aca Sci USA} 95: 4970-4975.

\bibitem{Ben}
Frauenfelder H and McMahon B (1998) Dynamics and function of proteins: The search for general concepts. 
{\it Proc Natl Aca Sci USA} 95: 4795-4797.

\bibitem{Aflalo}
Vakser IA and Aflalo C (1994) Hydrophobic docking: A proposed enhancement to molecular recognition techniques. {\it Proteins} 20:
320-329.

\bibitem{Young}
Young L, Jernigan RL and Covell DG (1994) A role for surface hydrophobicity in protein-protein recognition. {\it Prot Sci} 3: 717-729. 

\bibitem{Jones1997133}
Jones S and Thornton JM (1997) Prediction of protein-protein interaction sites using patch analysis. {\it J Mol Biol} 272: 133-143.

\bibitem{Tsai}
Tsai C-J, Lin SL, Wolfson HJ and Nussinov R (1997) Studies of protein-protein interfaces: A statistical analysis of the hydrophobic effect. {\it Prot Sci} 6: 53-64.

\bibitem{YBai07141995}
Bai Y, Sosnick TR, Mayne L and Englander SW (1995) Protein folding intermediates: native-state hydrogen exchange.
{\it Science} 269: 192-197.  

\bibitem{Fierz}
Fierzt B, Reiner A and Kiefhaber T (2009) Local conformational dynamics in $\alpha$-helices measured by fast triplet transfer.
{\it Proc Natl Aca Sci USA} 106: 1057-1062.  


       
\end{thebibliography}
\end{document}